# The application of spectrum standardization method for carbon analysis in coal using laser-induced breakdown spectroscopy

Xiongwei Li,[1] Zhe Wang[*,1], Yangting Fu,[1] Zheng Li,[1] Jianming Liu,[2] Weidou Ni[1]

[1] State Key Lab of Power Systems, Department of Thermal Engineering, Tsinghua-BP Clean Energy Center, Tsinghua University, Beijing, China

[2] China Guodian Science and Technology Research Institute, Nanjing, 100034, China

**Abstract:** Measurements of carbon content in coal using laser-induced breakdown spectroscopy (LIBS) is limited by its low measurement precision and accuracy. A spectrum standardization method was proposed to achieve both reproducible and accurate results for the quantitative analysis of carbon content in coal with LIBS. The proposed method utilized the molecular carbon emissions to compensate the diminution of atomic carbon emission caused by matrix effect. The compensated carbon line intensities were further converted into an assumed "standard state" with fixed plasma temperature, electron density, and total number density of elemental carbon, which is proportional to its concentration in the coal samples. In addition, in order to obtained better compensation for total carbon number density fluctuations, an iterative algorithm was applied, which is different from our previous standardization calculations. The modified spectrum standardization model was applied to the measurement of carbon content in 24 bituminous coal samples. The results demonstrated that the proposed method had superior performance over the generally applied normalization methods. The average relative standard deviation, the coefficient of determination, the root-mean-square error of prediction, and the average maximum relative error for the modified model were 3.44%, 0.83, 2.71%, and 12.61%, respectively, while the corresponding values for the normalization with segmental spectrum area were 6.00%, 0.75, 3.77%, and 15.40%, respectively, showing an overwhelming improvement.

## 1 Introduction

On-line analyses of coal properties are of thirst needs by power industry for coal pricing and combustion optimization [1-3]. As the most important index reflecting coal quality, carbon content in coal can provide quick estimation of the calorific value of coal and is very valuable for the operation of coal-fired power plants [4]. Laser-induced breakdown spectroscopy (LIBS) is a promising technique for on-line coal measurement with its capability of rapid analysis, minimal sample preparation, and simultaneous multi-elemental detection [5-7]. A number of studies on coal analysis by LIBS have been performed [8-13]. Among these studies, only a few studies focus on the quantitative determination of non-metallic elements (e.g., C, H, O, and N) in coal [11-13]. The measurement of these elements suffers from not only the low measurement accuracy and but also the low reproducibility due to multiple factors, such as matrix effects and variations in experimental conditions [14-16].

The most common method to improve LIBS measurement precision is the internal standard method, which is based on the calculation of the intensity ratio of the analyte and reference element [17,18]. To apply this method, a known or constant internal standard concentration is required, and the excitation potential of the reference element should be similar to that of the interested element. For the measurement of carbon content in coal, however, such an appropriate reference element cannot be found. Other methods that utilize background emission or spectral area to normalize the signal cannot effectively reduce the measurement uncertainty of carbon content in coal [19-22].

---
[*] Corresponding author. Tel: +86 10 62795739; fax: +86 10 62795736
  Email address: zhewang@tsinghua.edu.cn

We proposed in our previous study a spectrum standardization method by which the recorded characteristic line intensity was converted to line intensity at a "standard plasma state" with standard plasma temperature ($T_0$), electron number density ($n_{e0}$), and total number density of measured species ($n_{s0}$) [23]. A simplified spectrum standardization method was thereafter introduced to further improve the measurement precision and accuracy with much less calculation effort [24]. The application of the methods for brass alloy samples showed that the both spectrum standardization and simplified spectrum standardization methods improved both measurement precision and model accuracy and the later one yielded better results. However, the results were not satisfying when we directly applied the simplified spectrum standardization method to the measurement of carbon content in coal where exists complicated and strong matrix effects. In this paper, special measure was taken to modify the simplified spectrum standardization method for the measurement of carbon content in coal.

## 2 Method description

In this section, the simplified spectrum standardization model is briefly reviewed, and thereafter the modified spectrum standardization method is introduced.

### 2.1 Brief introduction of the simplified spectrum standardization model

In the spectrum standardization model, a "standard plasma state" was defined by characterizing the plasma with a set of constant parameters ($n_{s0}$, $T_0$, and $n_{e0}$), which were calculated in real application by averaging the corresponding plasma parameters of all the measurements [23]. With the assumed existence of the "standard plasma state", the deviation of the measured line intensity from the standard state intensity is caused by the fluctuations of plasma parameters ($T$, $n_e$, and $n_s$) from their standard state [24]. Using Taylor expansion,

$$I_{ij}(n_{s0}, T_0, n_{e0}) \approx I_{ij}(n_s, T, n_e) - (k_1 dn_s + k_2 C dT + k_3 C dn_e) \tag{1}$$

where $n_{s0}$, $T_0$, and $n_{e0}$ are the plasma parameters at the standard state, $I_{ij}$ is the raw measured line intensity, $I_{ij}(n_{s0}, T_0, n_{e0})$ is the calculated standard state line intensity, $C$ is the concentration of the specific element, and $k_1$, $k_2$, and $k_3$ are constants. Ignoring the self-absorption effects and the inter-element interference, $I_{ij}(n_{s0}, T_0, n_{e0})$ is proportional to the measured elemental concentration

$$I_{ij}(n_{s0}, T_0, n_{e0}) = k_0 C \tag{2}$$

The deviation of the $n_s$, $T$, and $n_e$ are also correlated to the measured spectral information. On the right hand side of Eq. (1), $dn_s$ is assumed to be proportional to the fluctuation in the sum of the multiple line intensities of the measured element. Considering that the excited states in plasma follow the Boltzmann distribution under the local thermodynamic equilibrium assumption, $dT$ may be associated with the intensity ratio of a pair of lines based on the principle of Boltzmann distribution [25]. The full width of half maximum (FWHM) of the spectral line is assumed to be proportional to electron number density, since the characteristic spectral line broadening is mainly caused by the Stark broadening for typical LIBS measurements. The deviation of electron density, $dn_e$, may be determined from the FWHM of the $H_\alpha$ spectral line through Stark broadening [25]. The $I_{ij}(n_{s0}, T_0, n_{e0})$ in Eq. (1) can be substituted by the concentration in Eq. (2). Then, from Eq. (1) we can obtain

$$C = a_1 I_{ij} + a_2 I_T + a_3 \left( \ln\left(\frac{I_2}{I_1}\right) - \left(\ln\left(\frac{I_2}{I_1}\right)\right)_0 \right) C + a_4 (\Delta\lambda_{stark} - (\Delta\lambda_{stark})_0) C + a_5 \tag{3}$$

In Eq. (3), $I_T$ is the sum of the multiple line intensities of the measured element. $a_1$, $a_2$, $a_3$, $a_4$, and $a_5$ are constants calculated from the regression process. Both $[\ln(I_2/I_1)]_0$ and $(\Delta\lambda_{stark})_0$ can be calculated from the all measured spectra's average which can be applied to indicate their standard state values. Rearranging Eq. (3) will give the concentration as

$$C = \frac{a_1 I_{ij} + a_2 I_T + a_5}{1 - a_3 \left( \ln\left(\frac{I_2}{I_1}\right) - \left(\ln\left(\frac{I_2}{I_1}\right)\right)_0 \right) - a_4 \left( \Delta\lambda_{stark} - (\Delta\lambda_{stark})_0 \right)} \quad (4)$$

The line intensity at the standard state is proportional to the measured elemental concentration. That is,

$$I_{ij}(n_{s0}, T_0, n_{e0}) = k_1 C = k_1 \frac{a_1 I_{ij} + a_2 I_T + a_5}{1 - a_3 \left( \ln\left(\frac{I_2}{I_1}\right) - \left(\ln\left(\frac{I_2}{I_1}\right)\right)_0 \right) - a_4 \left( \Delta\lambda_{stark} - (\Delta\lambda_{stark})_0 \right)} \quad (5)$$

The coefficients, $a_1$, $a_2$, $a_3$, $a_4$, and $a_5$, were calculated through the following two steps: 1) the starting values were calculated from the linear regression analysis of Eq. (3), in which the independent variable is a matrix composed of all measured values of $I_{ij}$, $I_T$, $\left( \ln\left(\frac{I_2}{I_1}\right) - \left(\ln\left(\frac{I_2}{I_1}\right)\right)_0 \right) C$, and $\left( \Delta\lambda_{stark} - (\Delta\lambda_{stark})_0 \right) C$, and the dependent variable is a vector composed of the contents of the interested element; 2) these coefficients were then optimized near the starting values with the target of minimum relative standard deviation (RSD) of the calculated elemental contents.

As described in Eq. (3), the standard state line intensity in the standardization equation is substituted by the concentration based on linear correlation, which avoids the direct calculation of the standard state line intensity. This substitution requires that there is good linearity between the standardized line intensity and the concentration of the interested element. However, the linear relationship between the atomic carbon intensity and the carbon content may be suffered due to the strong matrix effect in coal. Therefore, the substitution of the standard state line intensity by the concentration may not be accurate enough for estimation, and optimization with the target of minimum relative standard deviation (RSD) near the regression values may not obtain the accurate coefficients of the standardization equation. Moreover, for coal application, there were only two eligible carbon lines and using the sum of these two lines to compensate for the fluctuation of total number density of carbon may not be good enough, making the estimation of starting value even worse and decreasing the applicability of the simplified standard spectrum method for coal samples. That is, the simplified spectrum standardization method needs to be modified for coal application.

## 2.2 Model modification for coal application

In the coal LIBS spectrum, there is molecular emission from $C_2$ detected, which indicated that part of carbon is disappeared in generating the elemental lines. $C_2$ can be formed either from direct ablation of the coal samples or from the recombination of the carbon atomics. The model utilized the molecular emission from $C_2$ to compensate the emission intensity of atomic carbon based on the following reasons: 1) the formation of molecular $C_2$ will cause the reduction of atomic carbon emission [26]; 2) the increased amount of the ablated molecular $C_2$ due to the matrix effect will reduce the amount of the ablated atomic carbon. Then, the compensated intensity of carbon is

$$I_{ij} = I_C + m I_{C2} \quad (6)$$

where $I_C$ is the emission intensity of atomic carbon and $I_{C2}$ is the emission intensity of molecular carbon; $m$ is the coefficients calculated from the calibration of the carbon content using both of $I_C$ and $I_{C2}$. That is,

$$C = n_1 I_C + n_2 I_{C2} + n_3 \quad (7)$$

where $n_1$, $n_2$, and $n_3$ are the coefficients calculated from the regression process.

The relationship of the compensated carbon intensity and the carbon content should be

$$C = n_1 I_{ij} + n_3 \quad (8)$$

Then, the coefficient *m* can be obtained from:
$$m = n_2 / n_1 \tag{9}$$

As the compensated carbon intensity is still affected by the fluctuations of plasma parameters ($n_s$, $T$, and $n_e$), the deviations of the plasma parameters from their values at the standard plasma state are utilized to improve the precision of the compensated carbon intensity. That is, the compensated carbon intensity composed of the measured intensities of the atomic and molecular carbon is converted to the compensated carbon intensity at the standard plasma state.

As there are only two available emission lines for carbon, the number density of carbon may not be well represented by the sum of its emission line intensities. In the proposed model, the linear combination of several segmental spectral areas is used to represent the ablated mass of emitters in the plasma. Assuming the stoichiometric ablation, the number density of carbon is proportional to the product of the segmental spectral areas and the carbon concentration. Then, the deviation of the $n_s$ is,

$$dn_s = n_s - n_{s0} = \sum_{i=1}^{k} k_{1i} I_{Ti} C + k_2 C \tag{10}$$

where $I_{Ti}$ is a segmental spectral area; $k_{1i}$ and $k_2$ are coefficients.

Then the standardization equation becomes,

$$I_{ij}(n_{s0}, T_0, n_{e0}) = I_{ij} + \sum_{i=1}^{k} b_{1i} I_{Ti} C + b_2 C + b_3 \left( \ln\left(\frac{I_2}{I_1}\right) - \left(\ln\left(\frac{I_2}{I_1}\right)\right)_0 \right) C + b_4 \left( \Delta\lambda_{stark} - (\Delta\lambda_{stark})_0 \right) C \tag{11}$$

where $I_{ij}$ is the compensated carbon intensity; $b_1$, $b_2$, $b_3$, $b_4$, and $b_5$ are the coefficients calculated from the regression process, which will be discussed below.

If the self-absorption effect can be neglected, the linear function is chosen as the calibration function

$$C = k I_{ij}(n_{s0}, T_0, n_{e0}) + b \tag{12}$$

where $k$ and $b$ are constants calculated from the regression process, $I_{ij}(n_{s0}, T_0, n_{e0})$ is the standard state line intensity of the interested element obtained from Eq.(11).

From Eq. (6), Eq. (11) and Eq. (12), the concentration will be

$$C = \frac{k I_C + k m I_{C2} + b}{1 - k \left( \sum_{i=1}^{k} b_{1i} I_{Ti} + b_2 + b_3 \left( \ln\left(\frac{I_2}{I_1}\right) - \left(\ln\left(\frac{I_2}{I_1}\right)\right)_0 \right) + b_4 \left( \Delta\lambda_{stark} - (\Delta\lambda_{stark})_0 \right) \right)} \tag{13}$$

To perform the linear regression of Eq. (11), it requires the knowledge of the standard state line intensity. The standard state line intensity is initially estimated as the average value of repeatedly measured line intensities to obtain the coefficients in the standardization equation. The more accurate the standard state line intensity is estimated, the more accurate coefficients in the standardization equation can be calculated, and vice versa. Therefore, a new iterative procedure is added to the standardization equation to obtain a more accurate the standard state line intensity. The accuracy of the standard state line intensity is judged by the linear relationship between the standard state line intensity and the concentration.

As shown in Fig. 1, the iterative procedure for establishing the spectrum standardization model is performed as follows: 1) For each sample, the average of the compensated carbon intensities obtained from the repeatedly measurements is regarded as the initial value of the standard state line intensity; 2) the coefficients in Eq. (11) are calculated from regression process using the values of the standard state line intensities; 3) the converted line intensity for each measurement is obtained from the right hand side of Eq. (11) using the obtained coefficients, and the average of the converted line intensities for the repeatedly measurements is regarded as the new value of the standard state line intensity; 4) the coefficient of determination ($R^2$) of calibration curve between the standard state line intensities and the concentrations is calculated; 5) Steps 2) to 4) are repeated until the coefficient of determination ($R^2$) of calibration curve achieve the maximum, then the coefficients in Eq. (11) calculated from the last step are determined as the final values.

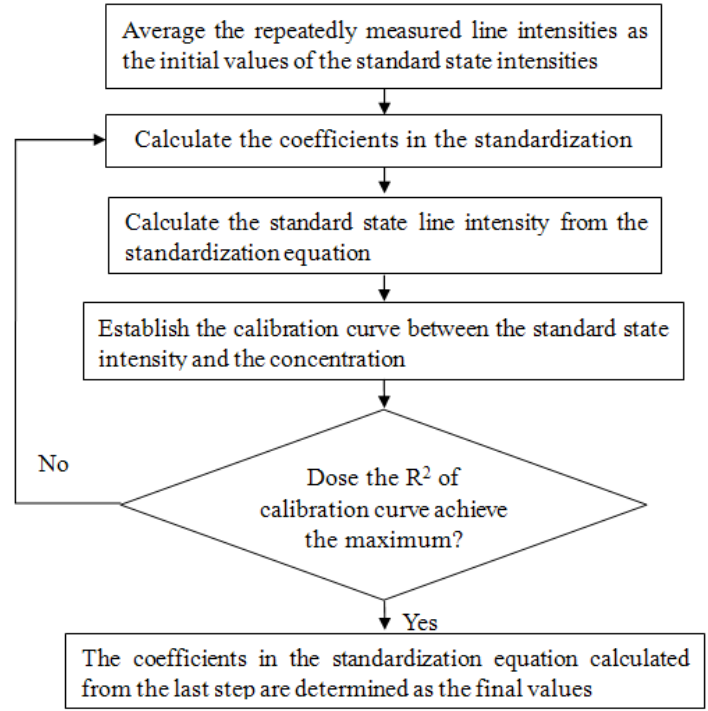

Figure 1. Flow diagram of the proposed iterative algorithm for the modified spectrum standardization method.

In this study, the analyte is carbon in bituminous coal. On the right hand side of Eq. (11), $I_{ij}$ is the combination of the integrated intensity of an atomic carbon line (193.029 or 247.856 nm) and the integrated intensity of molecular carbon $C_2$ in the range of 470-473.7 nm. On the right hand side of Eq. (11), $I_{Ti}$ (i=1, 2, 3) are the segmental spectral areas in the range of 190-310 nm, 310-560 nm, and 560-770 nm, respectively. The intensity ratio of two silicon atomic lines (212.412 and 250.689 nm) is used to monitor the plasma temperature variations. The Stark broadening terms are determined by the FWHM of the $H_\alpha$ spectral line after the Lorentz curve fitting.

## 3 Experimental setup

The LIBS system Spectrolaser 4000 (XRF Scientific, Australia) was used in this experiment. The experimental arrangement is similar to the one described previously [24]. Briefly, a Q-switched Nd:YAG laser emitting at 532 nm with a pulse duration of 5 ns was used as the ablation source. The laser energy was adjusted to be 120 mJ/pulse. The laser beam was focused onto the sample surface to create a plasma by a plano-convex quartz lens with a 150 mm focal length. The plasma emission was collected by four fiber optics, each directed to a Czerny-Turner spectrograph and detected by a charge coupled device (CCD). The four spectrometers and CCD detectors covered an overall range (nm) from 190 to 310, 310 to 560, 560 to 770, and 770 to 950, respectively, with a nominal resolution of 0.09 nm. The gate delay time was adjusted to be 2 μs, and the integration time was fixed at 1 ms.

The samples were 24 standard bituminous coals, which were certified by the China Coal Research Institute. Carbon was the interested element, and its concentration ranged from 42% to 82% (Table 1). Table 1 also gave the content of volatile matter in each coal sample on the air dry basis. The powder of each coal sample were placed into a small aluminum pellet die ($\phi$=30 mm, $h$=3 mm) which was pressed with a hydraulic jack under a pressure of 20 tons. The samples were mounted on an auto-controlled X-Y translation stage and exposed to air at atmospheric pressure. Similar to our previous study, the samples were divided into the calibration and validation sets. The calibration set provided spectral data for modeling, and the validation data set verified the accuracy of the model. To ensure a wide range and even concentration distribution in both sets, all samples were first arranged by their C concentrations, and then one of every three samples was chosen for validation.

Table 1. Carbon concentrations of 24 coal samples.

|  |  | 1 | 2 | 3 | 4 | 5 | 6 | 7 | 8 |
|---|---|---|---|---|---|---|---|---|---|
| **Calibration Set** | No. | 1 | 2 | 3 | 4 | 5 | 6 | 7 | 8 |
|  | C (%) | 47.12 | 52.61 | 53.77 | 54.72 | 58.12 | 59.84 | 67.18 | 67.77 |
|  | Volatile matter (%) | 11.31 | 23.23 | 14.03 | 13.1 | 30.43 | 28.65 | 18.21 | 34.46 |
|  | No. | 9 | 10 | 11 | 12 | 13 | 14 | 15 | 16 |
|  | C (%) | 70.45 | 74.7 | 76.69 | 77.28 | 78.64 | 79.02 | 79.98 | 81.54 |
|  | Volatile matter (%) | 14.41 | 33.4 | 33.41 | 32.22 | 33.9 | 11.42 | 31.92 | 12.43 |
| **Validation Set** | No. | 17 | 18 | 19 | 20 | 21 | 22 | 23 | 24 |
|  | C (%) | 53.42 | 55.67 | 59.91 | 72.71 | 75.96 | 78.58 | 79.7 | 81.45 |
|  | Volatile matter (%) | 25.58 | 19.11 | 28.9 | 30.91 | 32.94 | 32.41 | 15.3 | 11 |

Twenty-five locations were probed for each pellet. Each location was fired twice, i.e., first shot of 150 mJ to remove any contaminant, and the second shot of 120 mJ for analysis. A fan was used to blow off the aerosol particles after each laser shot to prevent signal change caused by aerosol production. All spectra were background-subtracted to reduce the systematic signal fluctuation. The intensity was defined as the integration of channel readings of an emission line above the background continuum. The system was warmed up for at least 1 h to ensure the thermal stability of the instruments.

# 4 Results and discussion

The performance of the modified spectrum standardization model is evaluated by comparing with the univariate calibration model with segmental normalization [27]. Four parameters are chosen to evaluate the performance of the modified spectrum standardization model. These parameters include the RSD of the spectral line intensity, the $R^2$ of calibration curve, the root mean square error of prediction (RMSEP) of mass concentration, the maximum relative error (MRE) of predicted mass concentrations. The RSD can evaluate the precision of measurement. The smaller the RSD is, the more precise the LIBS measurements will be. The $R^2$ can assess the quality of the data points that are used to establish the calibration model. The RMSEP and MRE can indicate the accuracy of predictions by the models.

## 4.1 Emission intensity of atomic and molecular carbon

As shown in Fig. 2 (a), the calibration between the spectral intensity of C(I) 247.856 nm and the carbon concentration for the 24 bituminous coals is poor, and the $R^2$ is 0.46. If only the coal samples whose volatile matter are lower than 23% are chosen for calibration, the $R^2$ will be increased to 0.90, as shown in Fig. 2 (b). The volatile in coal is a mixture of short and long chain hydrocarbons, aromatic hydrocarbons and some sulfur, and it will be liberated at high temperature. It shows that if the carbon content in two coal samples are almost the same, the emission intensity of atomic carbon in the high volatile coal sample will be obviously lower than that in the low volatile coal sample (e.g. No.7 and No. 8).

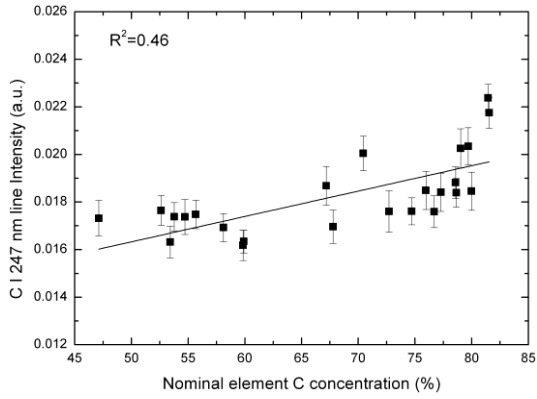 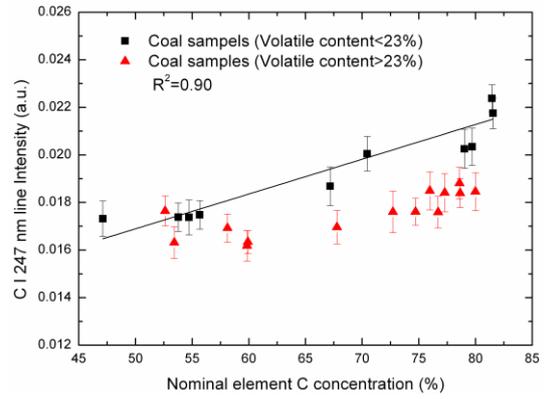

(a) Calibration for 24 coal samples  (b) Calibration for coal samples whose volatile content is lower than 23%

Figure 2. Calibration plots of the C (I) 247 nm line

Fig. 3 (a) shows the calibration plot between the emission intensity of $C_2$ in the range of 470-473.7 nm and the carbon concentration for the 24 bituminous coals, and the $R^2$ is 0.61. As shown in Fig. 3(b), if only the coal samples whose volatile matter are lower than 23% are chosen for calibration, the $R^2$ is 0.64; if only the coal samples whose volatile matter are larger than 23% are chosen for calibration, the $R^2$ is 0.74. It shows a contrary tendency for the emission intensity of $C_2$ compared with the spectral intensity of C(I) 247.856 nm. That is, if the carbon content in two coal samples are almost the same, the emission intensity of $C_2$ in the high volatile sample will be obviously larger than that in the low volatile sample (e.g. No.7 and No. 8).

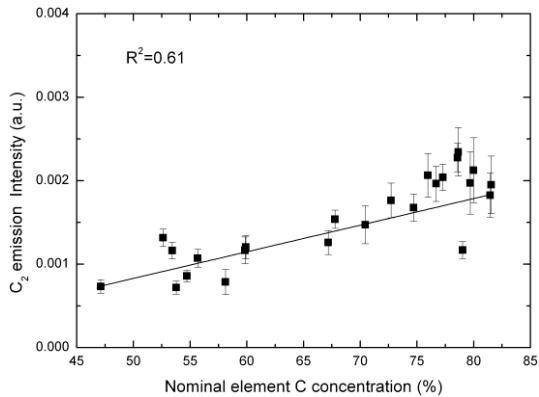 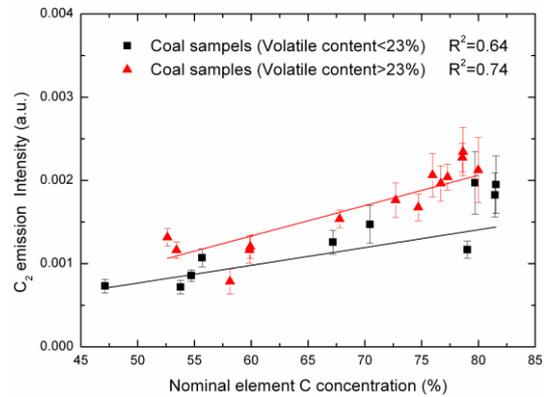

(a) Calibration for 24 coal samples  (b) Calibration for coal samples whose volatile content is lower than 23%

Figure 3. Calibration plots of the C (I) 247 nm line

According to the opposite tendency of the spectral intensity of C(I) 247.856 nm and the emission intensity of $C_2$, the molecular emission from $C_2$ was utilized to compensate the emission intensity of atomic carbon. The compensated carbon intensity can be obtained from Eq. (6) to Eq. (9).

## 4.2 Uncertainty reduction

The integrated intensity of C(I) 247.856 nm after segmental normalization was selected to establish the univariate calibration model. The integrated intensity of C(I) 247.856 nm and the integrated intensity of $C_2$ in the range of 470-473.7 nm were utilized to obtain the compensated carbon intensity, which was then utilized to establish the spectrum standardization model. The average RSD of the raw C(I) 247.856 nm line intensity is 3.79%, and the value of C(I) 247.856 nm after segmental normalization is 6.00%. It indicates that the segmental normalization method cannot effectively reduce the signal uncertainty.

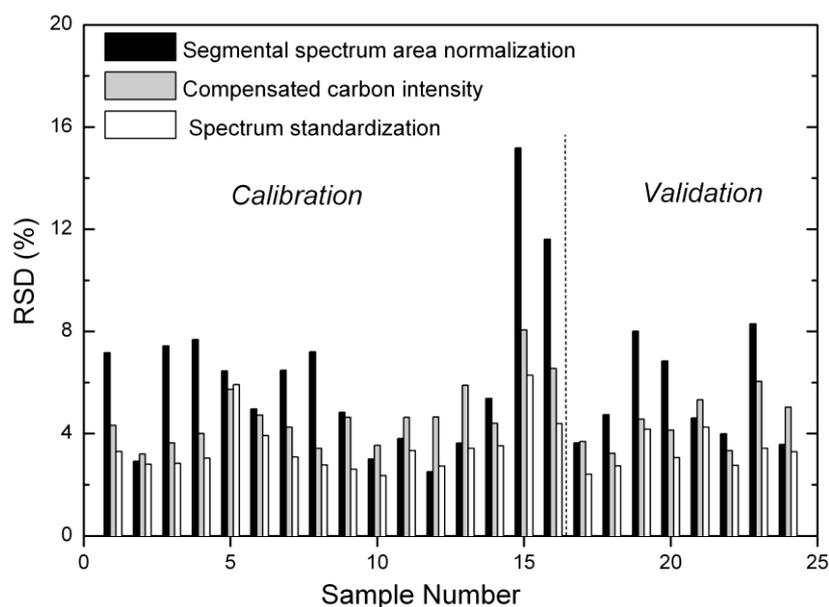

Figure 4. RSD values of the C(I) 247 nm line intensity for 24 coal samples from different spectral processing methods.

Figure 4 shows the comparison of the RSD values of C(I) 247.856 nm line intensity from different data processing methods for all the 24 bituminous coal samples. The average RSD of the compensated carbon intensity is 4.63%, and the value of the compensated carbon intensity with the modified spectrum standardization is 3.44%. The modified spectrum standardization method shows the best reproducibility because it not only compensate the diminution of the atomic carbon intensities caused by matrix effect, but also effectively compensate for pulse-to-pulse signal fluctuations caused by the variations of plasma parameters (plasma temperature, electron number density, and total number density of the measured element).

## 4.3 Accuracy improvement

The predicted carbon concentration of a sample is defined as the average value calculated from the 25 repeated measurements. The matrix effects of coal are so strong that the C(I) 247 nm line intensity does not have a good linearity with the carbon concentration (see Fig. 2(a)). The $R^2$ value of the calibration curve between the C(I) 247 nm line intensities and the carbon concentration is only 0.46. When the C(I) 247 nm line intensity was normalized with the segmental spectral area, the $R^2$ value of the calibration curve increased to 0.75 (Fig. 5). Given that the segmental spectrum area may linearly related to the ablation mass, the segmental normalization method can partly compensate for the variation in the ablation mass. Therefore, the segmental normalization method can improve the $R^2$ value of the calibration curve.

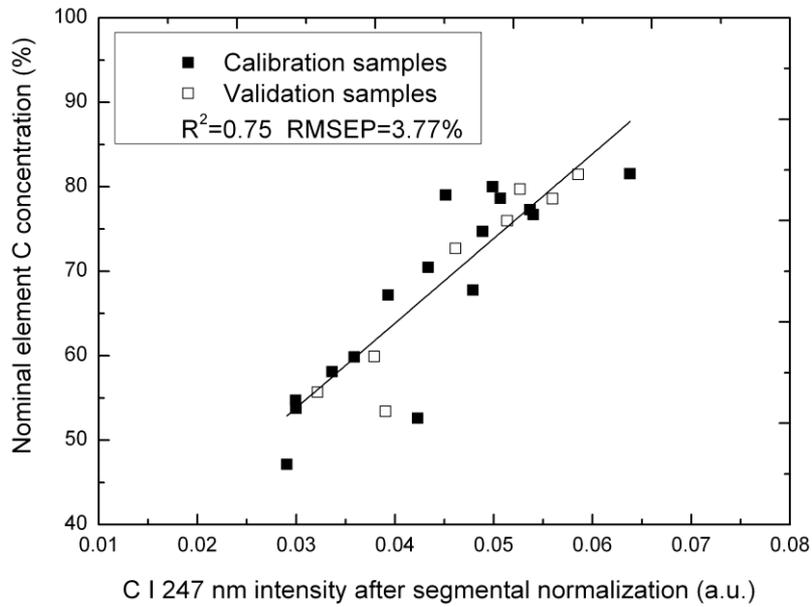
Figure 5. Calibration plots of the C(I) 247 nm line after segmental spectral area normalization

Fig. 6 shows the calibration plots of the compensated carbon intensity and the carbon content. The $R^2$ value of the calibration curve is 0.80. The performance of the model in prediction accuracy can be justified by the RMSEP. The RMSEP of the univariate model that utilizes the compensated carbon intensity is 3.02%, whereas the corresponding value for the univariate model with the segmental normalization is 3.77%. Compared with the segmental spectral area normalization, the improvement in $R^2$ and the reduction in RMSEP indicate that the emission intensity of $C_2$ can better compensate the diminution of atomic carbon emission caused by matrix effect.

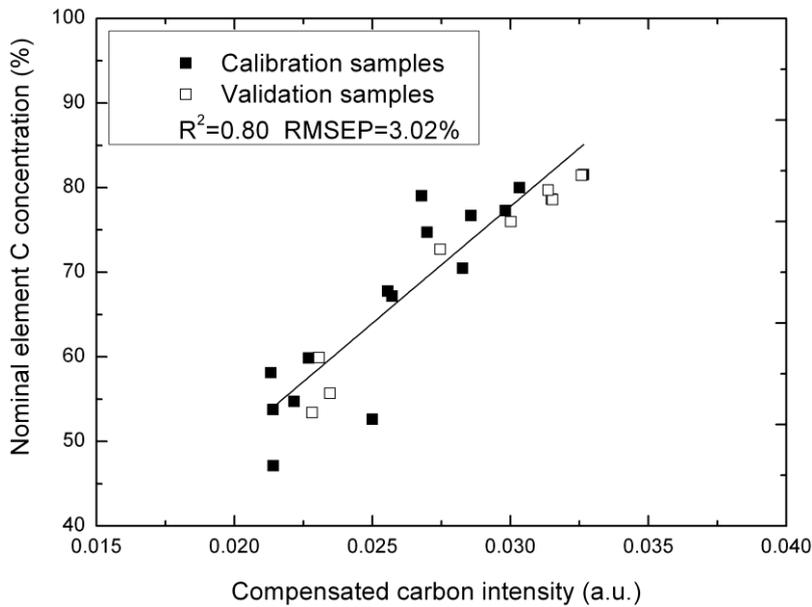
Figure 6. Calibration plots of the compensated carbon intensity

Fig. 7 demonstrates the calibration and validation results of the modified spectrum standardization model. The $R^2$ of the present model is 0.83 and RMSEP is 2.71%. Compared with the univariate model that utilizes the compensated carbon intensity, the increased $R^2$ indicates that the matrix effect is further corrected by compensating for the spectroscopic signal fluctuations attributed to the variation of plasma properties. The lowered RMSEP of the present model implies that plasma parameter correction terms ($T$, $n_e$, and $n_s$) in the present model contribute to the improvement in the prediction accuracy.

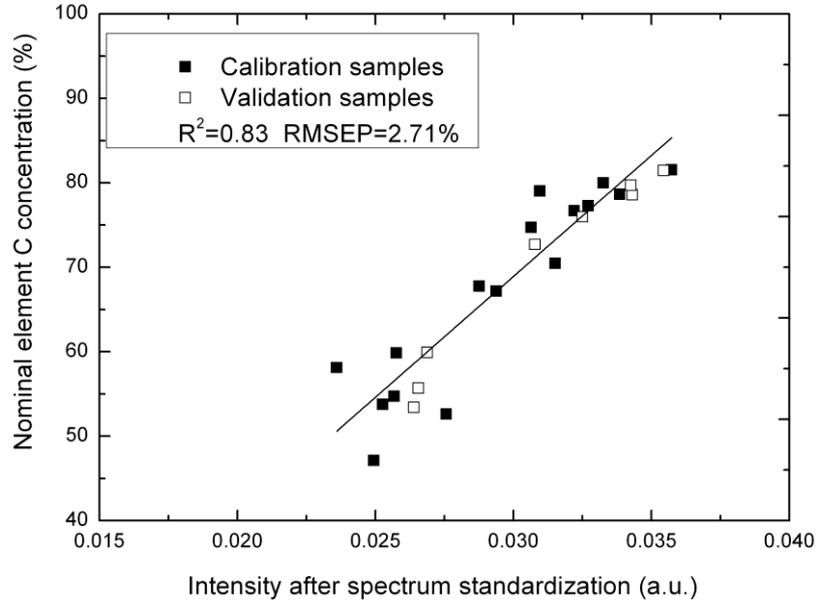
Figure 7. Calibration and validation plots for the spectrum standardization model.

The performance of the model in prediction accuracy can also be justified by MRE, which is defined as

$$\text{MRE} = \max\left(\left|\frac{C_{pre} - C_{nom}}{C_{nom}}\right| \times 100\%\right) \qquad (8)$$

where $C_{pre}$ is the predicted concentration for each measurement, and $C_{nom}$ is the nominal elemental concentration for the sample. The MRE indicates the largest deviation of a single measurement and can evaluate the prediction accuracy of the model especially for conditions where only single shot assay is allowed. The average MRE of the present model is 13.61%, whereas the value of the univariate calibration model with segmental spectral area normalization is 15.40%. These results further demonstrated the advantage of the present model in improving the measurement accuracy.

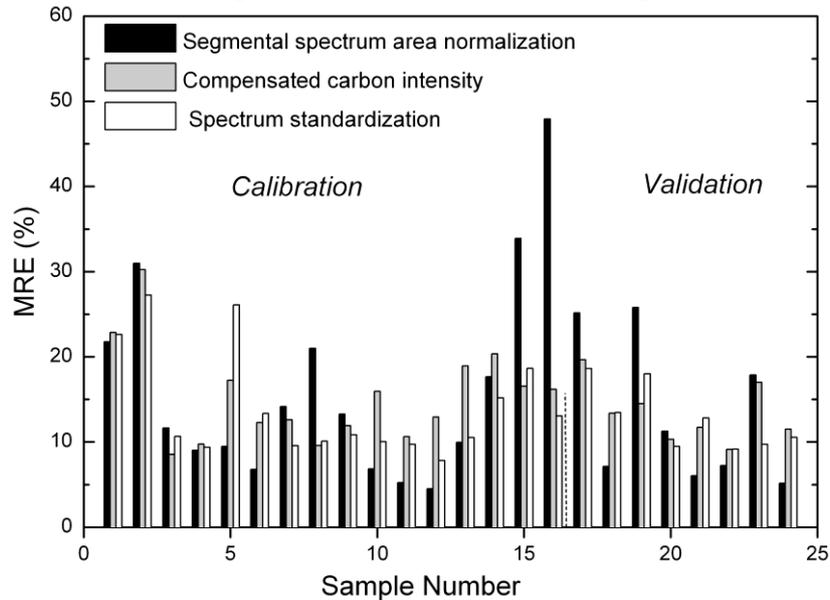
Figure 8. Maximum relative error of predicted C concentration for the different methods.

## 5 Conclusions

The previously reported simplified spectrum standardization method can obviously reduce the signal uncertainty and improve the measurement accuracy by converting the line intensity to the intensity at the standard plasma state. However, strong matrix effects of coal lead to an unsatisfactory linear relationship between the emission intensity and the concentration. Consequently, the simplified spectrum standardization method cannot be directly applied to the measurement of carbon content in coal. In the modified spectrum standardization model, diminution of the emission intensity of atomic carbon emission caused by matrix effect was compensated by utilizing the emission intensity of $C_2$. The modified spectrum standardization method also utilized the segmental spectral area to correct the fluctuations of ablated mass and employed an iterative algorithm to obtain more accurate standard state line intensities. The assay of the carbon concentration of 24 bituminous coal samples by the proposed model shows an improvement in both measurement precision and accuracy compared with the traditional single-variable method with segmental spectrum area normalization.

# Acknowledgement

The authors are grateful for the financial support from the National Natural Science Foundation of China (Grant No. 51276100) and National Basic Research Program (973 Program) (NO. 2013CB228501).

# References

[1] T. Ctvrtnickova, M. P. Mateo, A. Yanez, G. Nicolas, Appl. Surf. Sci. 257 (2011) 5447-5451.
[2] M. Kurihara, K. Ikeda, Y. Izawa, Y. Deguchi, H. Tarui, Appl. Optics 42 (2003) 6159-6165.
[3] T. Ctvrtnickova, M. P. Mateo, A. Yanez, G. Nicolas, Spectrochim. Acta B 65 ( 2010) 734-737.
[4] T. Yuan, Z. Wang, S. Lui, Y. Fu, Z. Li, J. Liu, W. Ni. J. Anal. At. Spectrom. 28 (2013) 1045-1053.
[5] D. W. Hahn, N. Omenetto, Appl. Spectrosc. 66 (2012) 347-419.
[6] D.A. Cremers, R.C. Chinni, Appl. Spectrosc. Rev. 44 (2009) 457-506.
[7] J. Yu, R. Zheng, Front. Phys. 7 (2012) 647-648.
[8] F. J. Wallis, B. L. Chadwick, R. J. S. Morrison, Appl. Spectrosc. 54 (2000) 1231-1235.
[9] T. Ctvrtnickova, M. P. Mateo, A. Yanez, G. Nicolas, Spectrochim. Acta B 64 (2009) 1093-1097.
[10] L. G. Blevins, C. R. Shaddix, S. M. Sickafoose, P. M. Walsh, Appl. Opt. 42 (2003) 6107-6118.
[11] L. Zhang, L. Dong, H. Dou, W. Yin, S. Jia, Appl. Spectrosc. 62 (2008) 458-463.
[12] Z. Wang, T. Yuan, S. Lui, Z. Hou, X. Li, Z. Li, W. Ni, Front. Phys. 7(2012) 708-713.
[13] S. Yao, J. Lu, J. Zheng, M. Dong, J. Anal. At. Spectrom. 27 (2012) 473-478.
[14] J. Li, J. Lu, Z. Lin, S. Gong, C. Xie, L. Chang, L. Yang, P. Li, Optics & Laser Technology, 41(2009) 907-913.
[15] B.C. Castle, K. Talabardon, B.W. Smith, J.D. Winefordner, Appl. Spectrosc. 52 (1998) 649–657.
[16] X. Li, W. Zhou, Z. Cui, Front. Phys., 7 (2012) 721-727.
[17] J.H. Kwak, C. Lenth, C. Salb, E.J. Ko, K.W. Kim, K. Park, Spectrochim Acta B, 64 (2009) 1105–1110.
[18] D. Mukherjee, A. Rai, M.R. Zachariah, Aerosol Sci., 37 (2006) 677–695.
[19] L. Xu, V. Bulatov, V.V. Gridin, I. Schechter, Anal. Chem., 69 (1997) 2103–2108.
[20] F.J. Fortes, M. Cortés, M.D. Simón, L.M. Cabalín, J.J. Laserna, Anal. Chim. Acta, 554 (2005) 136–143.
[21] D. Body, B.L. Chadwick, Acta Part B, 56 (2001) 725–736.
[22] J.A. Bolger, Appl. Spectrosc., 54 (2000) 181–189.
[23] Z. Wang, L.Z. Li, L. West, Z. Li, W.D. Ni, Spectrochim. Acta Part B 68 (2012) 58-64.
[24] L. Li, Z. Wang, T. Yuan, Z. Hou, Z. Li, W. Ni, J. Anal. At. Spectrom. 26 (2011) 2274-2280.
[25] V.P. Andrzej W. Miziolek, Israel Schechter, Cambridge University Press, Cambridge, UK, 2006.
[26] M. Dong, X. Mao, J. J. Gonzalez, J. Lu, R. E. Russo, Anal. Chem. 85 (2013) 2899-2906.
[27] J. Feng, Z. Wang, L. West, Z. Li, W.D. Ni, Anal. Bioanal. Chem. 400 (2011) 3261-3271.